%% file: aaskaii_IPS.tex
\documentclass[a4paper,11pt]{article}
\usepackage{aaskaiid}
\usepackage{orcidlink}
\usepackage{booktabs}

\newcommand\arcsec{\ensuremath{^{\prime\prime}}}

\title{Understanding sub-arcsecond scale radio sources and space weather with the SKA-Low using interplanetary scintillation}
\ShortTitle{IPS with SKA-Low}

\author[1]{Rajan Chhetri\orcidlink{0000-0002-6966-9357}}
\ShortName{Rajan Chhetri et al.} 
\emailAdd{rajan.chhetri@csiro.au} 
\author[1]{John Morgan\orcidlink{0000-0001-9224-5483}}
\author[2]{Harsha Avinash Tanti\orcidlink{0000-0003-4458-0679}}
\author[1,3]{Hayley Bignall\orcidlink{0000-0001-6247-3071}}
\author[4,5,6]{Emily Kerrison\orcidlink{0000-0002-0011-6922}}
\affiliation[1]{CSIRO, Space and Astronomy, P.O. Box 1130, Bentley, WA 6102, Australia}
\affiliation[2]{Department of Astronomy, Astrophysics and Space Engineering, Indian Institute of Technology Indore, Khandwa Road, Simrol, Indore 453552, Madhya Pradesh, India}
\affiliation[3]{Manly Astrophysics, 15/41-42 East Esplanade, Manly, NSW 2095, Australia}
\affiliation[4]{Sydney Institute for Astronomy, School of Physics A28, University of Sydney, NSW 2006, Australia}
\affiliation[5]{ARC Centre of Excellence for All Sky Astrophysics in 3 Dimensions (ASTRO 3D)}
\affiliation[6]{ATNF, CSIRO Space and Astronomy, PO Box 76, Epping, NSW 1710, Australia}
\author[2]{Abhirup Datta\orcidlink{0000-0002-5333-1095}}
\author[7]{Anshu Kumari\orcidlink{0000-0001-5742-9033}}
\affiliation[7]{Udaipur Solar Observatory, Physical Research Laboratory, Dewali, Badi Road, Udaipur - 313001, Rajasthan, India}

\abstract{
Continuum surveys with SKA-Low will provide excellent spectral information for detected objects. However, the angular resolution it provides cannot clearly distinguish between extended radio galaxies and their sub-arcsecond scale counterparts. We have previously used an interplanetary scintillation (IPS) technique with wide field-of-view instruments (the MWA and ASKAP) to detect and study $<$0.3\arcsec\ objects. The ability to identify hundreds of compact sources from large areas of the sky using $<$5 minutes of data, and associated low computational cost makes widefield IPS a very efficient approach to identify sub-arcsecond scale objects at low radio frequencies.

We aim to utilise SKA-Low's flexibility in data capture, frequency coverage and high sensitivity to develop widefield IPS technique with SKA-Low, enabling the identification of compact radio sources at angular scales more than an order of magnitude smaller than that achieved by SKA-Low's longest baselines. This capability will facilitate studies of compact source populations (e.g., AGN cores), their relationship to extended radio galaxies, AGN contributions in star-forming galaxies, identify candidates for pulsars and HI absorption, and provide a network of calibrators for direction-dependent ionospheric calibration for SKA-Low. Additionally, ground-based IPS observations are one of the very few remote sensing techniques to probe the Heliosphere which will enable SKA-Low to provide unique insights into the solar wind.

Here, we present recent developments in the field of compact sources at low frequencies, their synergies with other fields of astronomy and space weather, and outline the technical developments and approaches that are necessary with the SKA-Low to realise these studies.
}

\begin{document}
\maketitle
\include{journal-names}

\section{Introduction}
Interplanetary Scintillation (IPS) is a phenomenon that can be exploited to advance diverse science goals from across astrophysics and heliophysics.
It consists of variations in source intensity on timescales $\sim$1\,s, typically seen close to the Sun, although below 100\,MHz it is observed across the sky at the $\sim$10\% level \citep[e.g.][]{1970MNRAS.150..141H}.
For astrophysics, the key feature is that only sources that are sufficiently small can show interplanetary scintillation (the effect is slightly frequency dependent, but the critical size is $<1$\arcsec). For SKA-low, this means that the spatial scales probed by IPS exceeds the interferometric resolution due to its longest baselines (6.4 arcsec at 150 MHz) by as much as an order of magnitude, without the huge calibration and computational challenges incurred by low-frequency Very Long Baseline Interferometry (VLBI). IPS has been used in the past to study source structure (till the advent of VLBI) and continue to be used for space weather studies. 

The key feature of IPS for space weather studies, which is shared to some extent with other ground-based radio techniques, is the ability to remotely sense the Heliosphere.
IPS studies using SKA precursors (particularly the Murchison Widefield Array \citep[MWA;][]{2013PASA...30....7T} and the Low-Frequency Array \citep[LOFAR;][]{2013A&A...556A...2V}, but also Australian SKA Pathfinder \cite[ASKAP;][]{Hotan2021} have led to a remarkably wide range of science. The SKA-Low will allow us to go much further, primarily due much greater sensitivity.

This chapter is organised as follows: in Sect.~\ref{sec:literature} we summarise the wealth of science results which have sprung from the use of the IPS technique, with a strong emphasis on the work that utilized SKA precursors.
In Sect.~\ref{sec:ska} we describe how IPS observations will be made with the SKAO telescopes,
and in Sect.~\ref{sec:summary} we summarise our conclusions.

\section{Science with IPS}
\label{sec:literature}
After its discovery \citep{Clarke:1964}, the initial use-case for IPS observations was for high-resolution astronomy \citep{Hewish1964}, and it led indirectly to the discovery both of pulsars by Jocelyn Bell \citep{1968Natur.217..709H}, and of millisecond pulsars \citep{2024JAHH...27..453R}.

However, the importance of IPS as a high-resolution technique waned with the introduction of VLBI and higher-frequency interferometers, and in recent years it has been used far more widely as a probe of the heliosphere; leading to fundamental discoveries well in advance of in situ measurements \citep[e.g.][]{1971A&A....10..310E,1980Natur.286..239C,1996Natur.379..429G};
and also the development of IPS as a powerful technique suitable for Space Weather forecasting \citep[see][for a recent review]{2020FrASS...7...76J}.

In recent years several new generation of Low-frequency instruments has been constructed; e.g. LOFAR, the MWA, Long Wavelength array \citep[LWA;][]{2013ITAP...61.2540E}.
The ability of these new telescopes to make IPS and other heliospheric measurements was recognised extremely early on \citep{2003EAEJA.....7234O,2010SoPh..265..293O}.
However, the first observation of IPS was serendipitous during standard nighttime observing \citep{2015ApJ...809L..12K}, which was later confirmed by LOFAR \citep{2016ApJ...828L...7F}, and this motivated consideration of the impact of IPS on other low-frequency science \citep{2015ApJ...814...27T}.

Less anticipated was the continued utility of the IPS technique for high-resolution radio astronomy.
Low-frequency VLBI continues to be extremely challenging, although International LOFAR has motivated huge technical strides in recent years \citep[e.g.][]{2022NatAs...6..350S}.
However, in advance of this, a series of studies using the MWA \citep{Morgan2017,Chhetri2018-IPS3, Chhetri2018-IPS2,2019MNRAS.483.1354S,Morgan2019} demonstrated the capabilities unlocked by the combination of the wide field-of-view and snapshot imaging fidelity, and the flexibility and computational efficiency of new imaging software such as WSClean \citep{2014MNRAS.444..606O}.
This early work led to further studies to connect IPS to more conventional VLBI \citep{2022MNRAS.509.2122J}, assisted in the characterisation of calibrators for International LOFAR \citep{2022A&A...658A...2J}, and culminated in the largest single release of a compact radio source catalogue to-date \citep{2022PASA...39...63M}. More recently, the flexibility of the technique has been demonstrated through IPS detections with ASKAP \citep{2023AdSpR..72.5361C}, although at higher frequency, which in turn allows probing the interplanetary space closer to the Sun, as well as opening up entirely new astrophysics use cases.

\begin{figure}[ht]
	\centering
	\includegraphics[width=0.8\columnwidth]{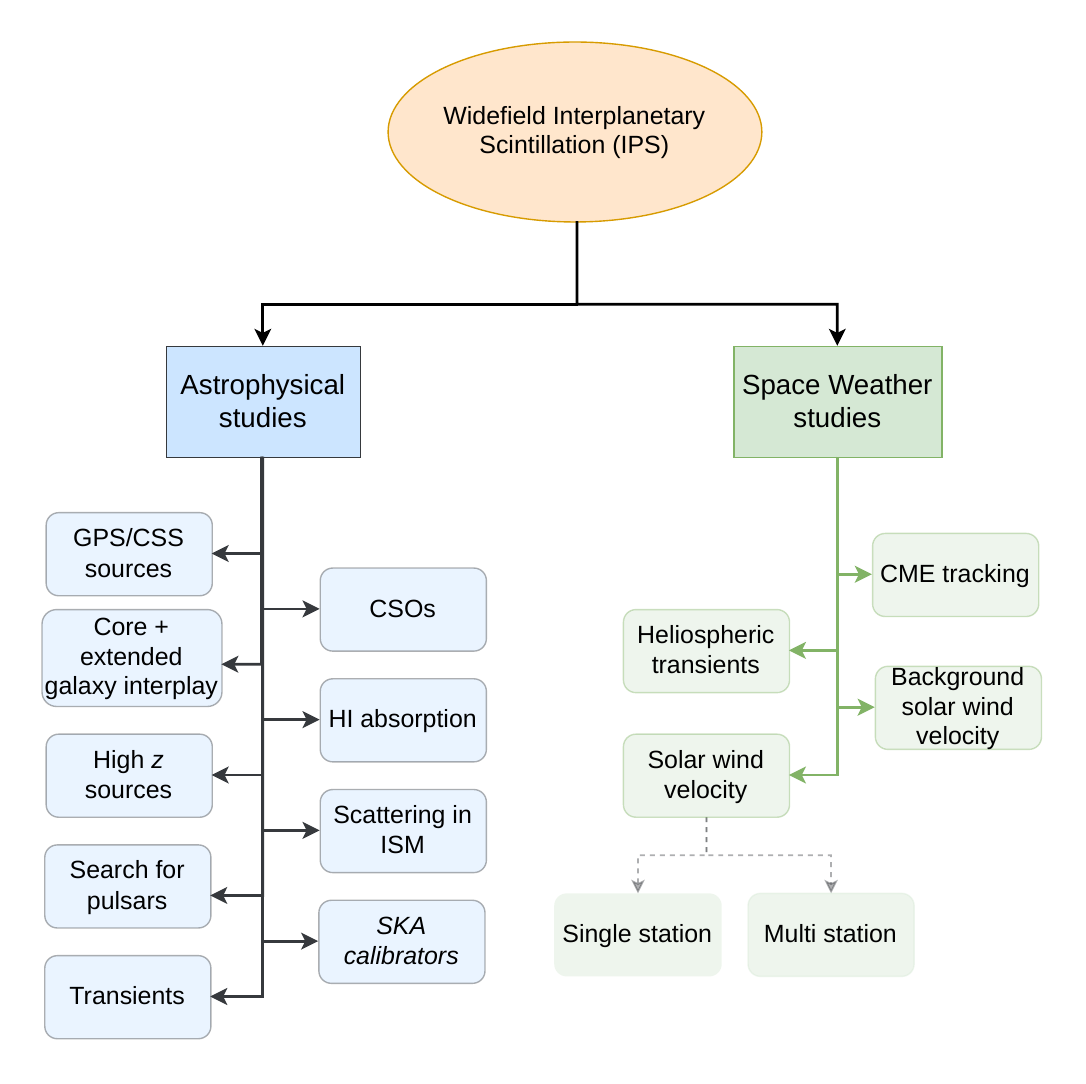}
	\caption{Summary of different science cases enabled by widefield IPS discussed in this section.}
	\label{fig:science_flowchart}
\end{figure}

We will now discuss these two distinct science objectives (astrophysical studies and space weather studies summarised in Figure \ref{fig:science_flowchart} in more detail.

\subsection{Astrophysics studies}
\label{subsec:astro_studies}

\subsubsection{GPS and CSS sources}

In the recent past, \cite{Chhetri2012-AT20G_Spec+zcutoff}, \cite{Chhetri2018-IPS2} and \cite{Chhetri2018-IPS3} identified that the population of sub-arcsecond scale sources changes significantly with survey frequency. They showed that at low frequencies, the population is dominated by sources with peaks in their spectra, with potentially up to 80\% of the sources' spectra being due to peaked spectrum sources. This includes both sources with peak in their spectra and compact steep-spectrum objects that are expected to be a more evolved form of the peak-spectrum objects (and have their peaks at very low frequencies, thus showing steep spectra). This is unlike at gigahertz frequencies where the population is mostly blazar type sources. At 800\,MHz, sample of sources which show a strong IPS signature is $\sim40\%$ and exhibit spectral peaks \citep{2026arXiv260712377G, 2026arXiv260628302C}. Inverting the way in which the GPS-IPS connection is studied, \citet{Kerrison2025} used IPS measurements for a sample of GPS/CSS sources peaking at a range of frequencies to analyse their sub-arcsecond structure, demonstrating the wider applicability of IPS to the extragalactic community. The sensitivity of SKA-Low will allow us to investigate how this population changes not only in frequency but also as a function of flux density. With new and upcoming surveys (e.g. Nancy Grace Roman Telescope \citep{2019arXiv190205569A}, Euclid \citep{2025A&A...697A...1E}, The Dark Energy Spectroscopic Instrument \citep[DESI,][]{2022AJ....164..207D})) to provide redshift information, it will be possible to understand the nature of this evolution more fundamentally. 

\subsubsection{Compact Symmetric Objects (CSOs)} 
The compact steep spectrum sources identified by IPS indicates that there is a small interesting subpopulation of CSS sources that are unresolved by IPS at both MWA and ASKAP frequencies (i.e. smaller than the characteristic angular scales probed by IPS at the frequencies of their operation $\sim$ 0.3 and 0.1 arcseconds, respectively). This is potentially the rare class of objects that are very compact ($<$1\,kpc scale) extragalactic sources that show symmetric morphology in VLBI images. It is unclear whether they are part of the evolutionary picture of radio galaxies or an entirely different class of objects \citep[e.g.][]{ 2024ApJ...961..242R}. Only a small number of bright CSOs have been confirmed \citep{Kiehlmann2024}. IPS with SKA-Low will detect candidate CSOs in large numbers, which can be confirmed with spectral and variability studies, thus,  enabling population studies to understand this mysterious class of objects and their relationship to radio galaxy evolution.    

\subsubsection{The relationship between compact and extended source populations}

In studying extragalactic radio sources the two most common methods for identifying compact AGN are using their broadband radio spectral energy distribution (SED), and through VLBI imaging. On the one hand, peaked spectrum (PS) and Compact Steep Spectrum (CSS) sources are two classes of compact (with linear size 0.5 to $\sim$ 20\,kpc) radio AGN identified on the basis of a broadband spectral peak and an inverted spectrum, respectively \citep{Odea2021}. They are typically thought to be either young ($\sim10^3\,$years post-triggering) or impeded in their growth by the dense, ambient medium of their host, these radio AGN offer a window into jet-ISM interactions, but samples are often contaminated by beamed blazars \citep[e.g.][]{Mingaliev2012}. On the other hand, CSOs are identified morphologically as sources with their flux distributed \textit{symmetrically} between two or three components spanning a few hundred parsec, in contrast to the \textit{asymmetric} distribution of core-jet radio AGN \citep{Readhead1993}. These two populations of compact sources identified spectrally and morphologically likely have some overlap, although the exact relationship they have between each other, and with the population of extended radio sources is still a subject of debate \citep[e.g.][]{Sadler2016, Kiehlmann2024}.

IPS observations of both PS and CSO samples will allow for a direct comparison between the compact, extragalactic radio sources selected via each criterion, providing a third measure of angular scale independent to the method use for source identification. More generally, widefield IPS observations capturing a diverse population of extragalactic radio sources will help to characterise the transition from compact to extended radio structure, with IPS observations at \textit{different frequencies} using both SKA-Low and -mid adding information on multiple spatial scales.

IPS patterns encode source structure on scales smaller than the resolution of SKA-Low, which can be modelled via IPS power spectra. An example of this for the case of interstellar scintillation (ISS; which is sensitive to smaller physical scales, but is essentially the same phenomenon) is clearly illustrated in \citet{2007MNRAS.380L..20M}.
By comparing power spectra at different epochs, the authors modelled the emergence of a new compact jet component in a beamed quasar. For that case, many hours of dedicated observation with the WSRT were required. Due to the much shorter timescales of IPS, power spectra can be estimated with significantly higher accuracy than generally possible for ISS, for many thousands of sources and at multiple frequencies from only minutes of observing with SKA-Low.  While CSOs would be largely unresolved by IPS, double radio sources with kiloparsec-scale separations (so-called intermediate symmetric objects) could be measured through oscillations in the IPS power spectra. Such sources represent an important evolutionary stage in radio galaxy formation, and comparing their numbers with those of CSOs and more extended sources in an unbiased way through IPS would be extremely useful.  

\subsubsection{HI absorption studies}
Probing neutral hydrogen (H\,I) across the Universe provides a means to study the evolution of galaxies. In radio frequencies, the 21\,cm hyperfine transitions in the hydrogen atom can be used to trace H\,I. The strength of H\,I in emission is dependent upon column density and so is its detection is limited by the sensitivity of the instrument. Absorption of H\,I is an excellent alternative as its strength is dependent upon the column density as well as the background radio continuum source against which the absorption is measured, and therefore independent of redshift. \cite{2024MNRAS.527.8511A} used data from the ASKAP First Large Absorption Survey in H\,I (FLASH) to search for H\,I absorption against bright sources from Molonglo Reference Catalogue 1-Jy sample, and found 3 absorbing samples (one of which was the strongest case of H\,I absorption). They found that 2 out of these 3 sources were strong IPS sources (the third source was outside the coverage of the MWA IPS survey area), and attribute their low detection rate (1.8\%) to the preponderence of large number of extended radio galaxies. Therefore, H\,I absorption candidates based on IPS (with prior knowledge of compactness) strongly increases the chances of detection of H\,I absorption. Similar very positive results have also been seen with ASKAP IPS and FLASH studies \citep{2026arXiv260712377G}. Hence, the compactness information provided by sensitive IPS studies from SKA-Low combined with the SKA H\,I absorption studies will form a very efficient pair to conduct absorption studies down to much lower brightness sources compared to currently possible, and will contribute strongly to studies of evolution of galaxies. 


\subsubsection{High redshift sources}
In the literature, searches for radio loud AGNs at high redshift (z$\gtrsim$5) have pre-supposed that their search efficiency are raised by avoiding extended radio galaxies. This is based on the assumption that these objects are expected to be young, luminous radio galaxies, and have often used peaked/curved source spectra as a proxy for angular size scales. \citet{2019MNRAS.483.1354S} found that median redshift of compact IPS sources in \cite{Chhetri2018-IPS2} have a redshift of $\sim$1.5, and that at least 30\% of them are likely to be at $z>2$. Using multi-wavelength data and angular size estimates from IPS, \cite{2021PASA...38...49D} argued that their candidate high redshift object to be at $z>7$. Similarly, \cite{2022PASA...39...61B} found that that their candidate high redshift sources detected using curved spectra and non-detection in \textit{k-band} were preferentially IPS sources. Therefore, IPS surveys, especially with the high sensitivity that the SKA-Low will provide, will contribute strongly to raise the efficiency of the search for elusive radio-loud high-redshift sources, helping us understand the interaction between the massive galaxies and their central supermassive black holes in the early Universe. 


\subsubsection{Measurement of Scattering by the Interstellar Medium}
IPS will be suppressed if the source size is larger than $\sim$0.3\arcsec at 150\,MHz.
This means that sources that are sufficiently scatter-broadened by the interstellar medium will not show IPS, or only at a reduced level.
IPS as a probe of interstellar scattering has a long history \citep{1972Natur.236..440R} with measurements of scattering within a few degrees of the Galactic Plane \citep{1984Natur.312..707R,1988MNRAS.234..853P}, as well as claims of the scattering across large regions of the sky due to the local bubble \citep{1992Natur.355..232H}.
Additionally, \citet[][see Figure~10]{2022PASA...39...63M} have demonstrated a lack of IPS sources towards H\textsc{ii} regions.

The SKA-Low will have the ability to probe this scattering in exquisite detail due to the high sensitivity, and therefore a very high density of sources.
Additionally, the scatter-broadened size of sources scales with $\lambda^{2.2}$ \citep[e.g.][]{1993ApJ...411..674T}, while the critical size for scintillation scales only with $\sqrt{\lambda}$ \citep[e.g.][]{1992RSPTA.341..151N}.
This means that the SKA-Low will be able to probe a very wide range of scattering regimes, but also, by making IPS observations at a broad range of frequencies, it will be possible to separate the effects of scatter broadening and source structure.

Finally, we note the synergy that such observations have with maps of Galactic Faraday rotation, which probe the Line-of-sight magnetic fields.
VLBI Measurements of stronger Galactic scattering \citep[e.g.][]{2018AJ....155..218K} show high levels of anisotropy, which is likely to probe the plane-of-sky magnetic field.
In principle, anisotropic scattering can be detected directly from IPS observations; however even if this turns out to be challenging, IPS can at least direct VLBI studies towards promising sources.

\subsubsection{Search for pulsars}
The SKAO telescopes will detect pulsars in unprecedented numbers for a wide range of science studies and applications. Conventional pulsar searches are multidimensional (sky location, dispersion measure, pulse period, and potential duty cycle / acceleration considerations) and therefore have a high computational demand \citep{Keane+2015aska.confE..40K}. Using image based techniques \citep[e.g.][]{Dai2016} to search for compact objects that show diffractive interstellar scintillation (DISS) is an interesting approach but is limited in detection volume as the characteristic bandwidth and timescale of DISS decreases strongly with distance. On the other hand, pulsars will show high IPS irrespective of their location in the galaxy, however candidate lists will also be contaminated by compact AGNs. \cite{Chhetri2018-IPS2} used IPS and steep spectral index ($\alpha\,<-0.7$) criteria to reduce this contamination by a factor of $\sim$45, strongly improving the practicality of conventional pulse search techniques. Additionally, image based IPS provides an efficient approach to identify exotic candidate pulsars such as those that have highly dispersed pulses (as well as potentially those in highly scattering regions) in the Galactic plane.

\subsubsection{A network of calibrators for the SKAO telescopes}
Blind IPS survey of the sky can provide a dense network of compact sources as demonstrated by \cite{Chhetri2018-IPS2} and \cite{2022PASA...39...63M}. IPS measurements of source size have been cross-validated with international LOFAR, with excellent agreement \citep{2022A&A...658A...2J}. Extrapolating the source counts of MWA detected IPS sources, \cite{Chhetri2018-IPS3} estimated 0.2 sources per square degree $<$0.3 arcsec in angular scale brighter than 0.4\,Jy (0.7 sources per square degree if sources with more complex sub-arcsecond structures are used). This further extrapolated to 4.6 sources per square degree at 0.04 Jy level, which would provide the necessary grid of calibrators important for direction dependent calibration of each isoplanatic patch in SKA-Low images.  

\subsubsection{Transients}

New instruments (e.g. Vera C. Rubin in optical wavelength and Cherenkov Telescope Array Observatory (CTAO) in high energy gamma ray frequencies) will be operational on similar timescales as the SKAO telescopes and will detect a large number of transients. Our IPS observing strategy (discussed in Section~\ref{Sec:Observing_strategy}) observes the sky in regular intervals and will detect compact radio transient sources. Therefore, IPS studies with SKA-Low is very timely for radio time domain astronomy, providing excellent synergy with multi-wavelength and multi-messenger studies. Please see chapters (e.g. \cite{Bignall01.2026.SKA} and the overview chapter for transient science \citep{Miller-Jones01.2026.SKA}) in this volume for detailed discussion of time domain studies.  

\subsection{Space Weather Studies}
\label{subsec:swx_studies}
Existing IPS facilities are listed in Table~\ref{tab:ips_instruments}.
While IPS Space Weather studies have typically relied on purpose-built instruments, LOFAR and MWA have demonstrated the utility of general-purpose low-frequency instruments.
Both are capable of making single-station measurements of IPS sources, but additionally, the MWA and International LOFAR have found their respective niches in widefield IPS with high densities of sources; and multi-station IPS with unprecedented number of baselines.

\begin{table}[htbp]
\centering
\small 
\caption{Major radio facilities currently used for IPS studies: ISEE \citep{1995JGG....47.1107A}; MWA \citep{2013PASA...30....7T}; LOFAR \citep{2013A&A...556A...2V}; Ooty \citep{2015ExA....39..319N}; MEXART \citep{2022RaSc...5707317G}; Pushchino \citep{2023AdSpR..72.5371C}.}
\label{tab:ips_instruments}
\begin{tabular}{@{}lll@{}}
\toprule
\textbf{Instrument}           & \textbf{Location}          & \textbf{Frequency} \\
\midrule
ISEE                          & Japan                      & 327\,MHz \\
MWA                           & Western Australia          & 80--300\,MHz \\
LOFAR                         & Netherlands \& Europe      & 10--250\,MHz \\
ORT                           & Ooty, India                & 327\,MHz \\
MEXART                        & Mexico                     & 140\,MHz \\
Pushchino BSA                 & Russia                     & 111\,MHz \\
SKA1-Low (Phase 1, $\sim$2028)& Western Australia          & 50--350\,MHz \\
\bottomrule
\end{tabular}
\end{table}

Although the SKAO instruments will have unique capabilities in this sphere, studies have also demonstrated the value of a world-wide network of IPS capable instruments for space weather studies \citep{2023AdSpR..72.5341J,2023AdSpR..72.5328I}.

\subsubsection{Space Weather studies using the widefield imaging approach}
Although the utility of the MWA's wide field of view for space weather measurements was clear from the first observations, the first study to demonstrate the utility of the MWA for CME tracking was that of \citet{2023SpWea..2103396M}, demonstrating for the first time that the unprecedented field of view and density of IPS sources, measured simultaneously in a single observation, allowed IPS to ``image'' a CME. SKA-Low, with its higher sensitivity will detect more compact sources than the MWA providing a higher definition to such "images" of CMEs.
This was followed by a study by \citet{Waszewski2023}, who showed that, with observations of a heliospheric transient spaced $\sim$1\,hour apart, the plane of sky velocity could be measured.
The nature of this transient, and the ability of the MWA to detect features invisible to other observations was clarified in \citet{2025arXiv251106670W}. 

Radio telescopes are also extremely well-suited to longer-term studies of larger scale structures, as shown by \citet{2025SpWea..2304342W}
Additionally, ASKAP can also be used for widefield IPS as demonstrated by \citet{2023AdSpR..72.5361C}, and the more recent paper describing IPS observations over the full field of view of ASKAP as a routine capability \citep{2026arXiv260628302C}. 

\subsubsection{Space weather studies in multi-station mode}
\label{subsubsec:swx_multistation}
By sampling the scintillation pattern at multiple points on the ground it is possible to determine the speed of the solar wind via the time delay between scintillation pattern observed at different locations, and the relative geometry of the solar wind velocity vector and the baseline.
Dedicated arrays designed specifically for this purpose have been the workhorses for IPS (for space weather purposes) over recent decades; firstly, the array operated in the US \citep[][defunct since the 1980s]{1972JGR....77.4602A}
and, continuously for the last 3 decades in Japan \citep[][still operational]{1995JGG....47.1107A}.

The ability for longer baselines to resolve multiple streams along the line of sight has been known for some time \citep{1996Ap&SS.243...87C,1996Natur.379..429G}.
However, it is pioneering work by the LOFAR telescope that has finally unlocked the enormous potential of a very large number of baselines, out to several times the Fresnel scale \citep{2022A&A...667A..57F,2023AdSpR..72.5311F}.
International LOFAR is an outstanding instrument in this respect, and novel approaches to presenting and interpreting these very rich data are still ongoing.

\subsubsection{Multi-frequency IPS}
In common with other scintillation phenomena, the modulation index in the weak scatter regime is proportional to observing wavelength.
This scattering strength depends on solar elongation (due to the decrease in the density of Heliosphere with increasing distance from the sun) as a power law of the sine of solar elongation \citep{1973JGR....78.1543R}\footnote{Additional variability is seen with the solar latitude from which the solar wind probed originated, as well as stage of the solar cycle, as well as temporal variation on shorter timescales due to space weather \citep{Waszewski2023}.}.

As explained in Section~\ref{Sec:Observing_strategy}, the principle consequence is that different solar elongations are appropriate for different observing frequencies.
Observing too close leads to observations in the strong regime, which can provide useful information, but require more detailed analysis. 
Observing too far away leads to a weaker scintillation signature and therefore a reduced signal-to-noise.

There are, additionally, observational techniques which rely on observing at different frequencies. \citet{1983A&A...123..191S} demonstrate the velocity can be determined by observing at two frequencies spaced by a factor of two.
Additionally, the report on a cross-frequency delay due to differing refraction of the ray paths due to the bulk reduction in density of the Heliosphere with increasing distance from the Sun.
This affect depends on the difference between the inverse wavelength squared at the two frequencies, and is therefore much stronger at lower frequencies. This should in principle provide an alternative route to velocity measurements, though the details remain untested.

\section{IPS with the SKAO}
\label{sec:ska}
In this section we describe how the SKAO telescopes would be used to observe IPS.
We have focused on SKA Low to capitalise on using its wide FoV, and because most IPS science with SKA precursors has utilised LOFAR and MWA. 
However, there is certainly value in using SKA-mid also, as it allows measurements to be made much closer to the Sun.

We envisage two potential observations modes for IPS.
The first, described in Sections~\ref{subsec:ska_imaging}\&\ref{subsec:ska_imaging_swx} is similar to the use of the MWA for IPS observations, using an inteferometric imaging approach.
The second, described in Section~\ref{subsec:ska_multi} is more similar to the LOFAR approach of beamforming individual stations (although a visibility-based approach is also possible).

\subsection{What is required to conduct IPS survey with the SKA-Low}
\label{subsec:ska_imaging}
The requirements for making survey the SKA-Low are simple: (i) correlation with $\sim$0.5 second time resolution in order to adequately sample the IPS signature (a shorter time constant would be desirable, and may even be necessary at high/frequencies or high solar wind speeds), (ii) our main approach is image based, therefore the sub-array selected should provide a good instantaneous UV coverage such that high quality images can be made at that time resolution.  

The current SKA-Low specifications provide visibilities at 0.85\,s time resolution for sub-arrays and 0.28 second for sub-arrays using sub-stations based on outputs of xGPU correlator. While the former may just be viable, it is far from ideal, and will average over IPS timescales of interest. The latter is a more useful visibility integration period for IPS studies. 

\begin{figure}[ht]
	\centering
	\includegraphics[width=0.4\columnwidth]{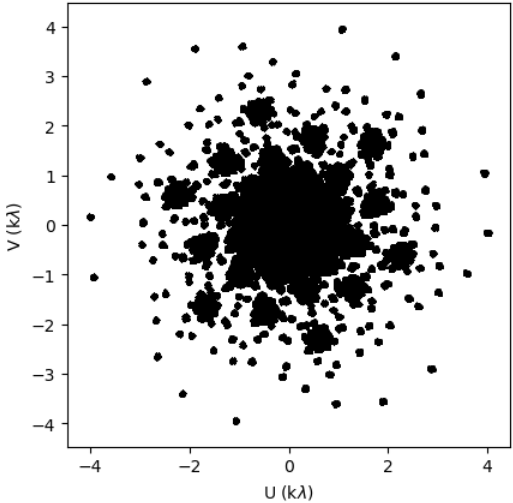}
	\caption{Instantaneous UV coverage of the SKA-Low (AA4) using only the stations in the inner 6 km of the array, with observing frequency centred on 162 MHz. }
	\label{fig:uv_coverage_162MHz}
\end{figure}

To meet the second requirement, stations including the core and the stations in the inner spiral arms (within $\sim$ 6\,km diameter with $\sim$20 stations) will be used. In Figure \ref{fig:uv_coverage_162MHz} we have plotted the instantaneous UV coverage of the SKA-Low at zenith using only the inner 6 km stations, with 32 MHz of bandwidth centred on 162 MHz (the frequency of our MWA surveys conducted with 16\,MHz bandwidth). Clearly, the UV plane is extremely well filled, which can be expected to provide excellent high time resolution images for IPS. The very highly filled UV plane will also ensure that the computation load on imaging is lowered as only light (if any) deconvolution will be required. Since IPS amplitude variations is proportional to wavelength, intend to limit our frequency bandwidth to such a value such that avoid any decorrelation that may happen to IPS signal at the two ends of the bandwidth. Then, either by observing at specific frequency ranges during observations or by limiting bandwidth during data processing we can use IPS across the SKA-Low bandwidth to study multiple interesting science cases such as: change in source structure with frequency, scatter broadening of compact sources as a function of Galactic latitude and multi frequency IPS. As an example, simultaneous IPS observations at evenly spaced frequency ranges 80--112\,MHz, 146--178\,MHz, 212--244\,MHz, 278--310\,MHz, selected to avoid known radio frequency interference frequencies, would strongly contribute to our science objectives.

\subsection{Strategy for covering large area of the sky}
\label{Sec:Observing_strategy}
\begin{figure}[ht]
	\centering
	\includegraphics[width=0.8\columnwidth]{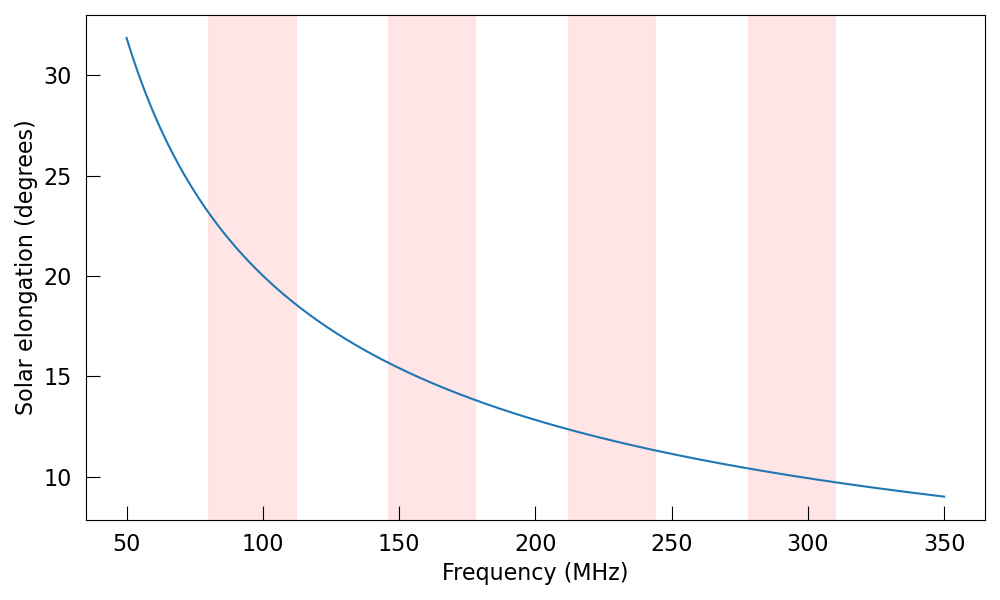}
	\caption{Solar elongations boundaries for strong and weak IPS regimes as a function of frequency. Solar elongations below the blue line indicate strong regime. This blue line provides an indication of how far from the Sun we want to place our beams to make our observations (above the blue line). The pink ranges indicate the four frequency ranges we envisage conducting our surveys in.}
	\label{fig:sol_el_range}
\end{figure}

The IPS survey would cover the sky in a similar manner to the strategy of following the Sun over the year, as done with the MWA and ASKAP. These observations would be made at solar elongations where IPS enters the weak regime as shown in Figure \ref{fig:sol_el_range}. 
For simplicity, to illustrate our strategy, we outline an observing plan using the 162 MHz observations only.

\begin{figure}[ht]
	\centering
	\includegraphics[width=0.8\columnwidth]{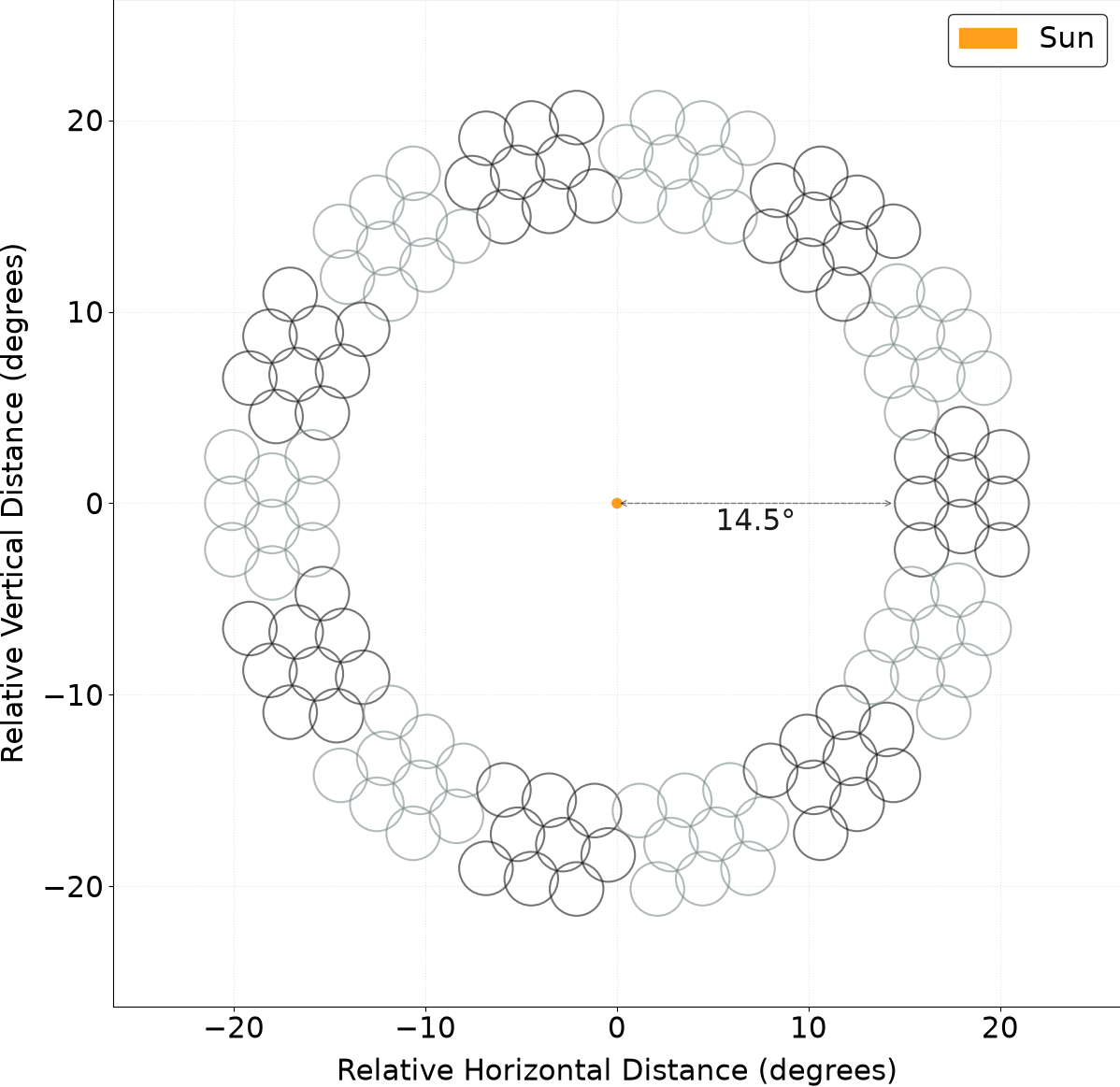}
	\caption{An example strategy of observing around the Sun to cover wide area with the SKA-Low. Each coloured closed-packed hexagonal grid of SKA-Low beams would be observed in 5 minutes covering $\sim$55 sq. deg. With 70 minutes of observations, this strategy would survey 770 sq. degrees in a day. Such observations would be repeated at a regular cadence (e.g. very 7 days) over the year to cover a large area around the Ecliptic.} 
	\label{fig:observing_grid}
\end{figure}

With only 32 MHz of the bandwidth needed for each observation, we can form 9 beams at any time. With FoV of 2.8 degrees at 162 MHz, by close packing them in 3$\times$3 layout, the footprint can cover up to 55 sq degrees of the sky with one pointing. With the inner edge of such a footprint pointed at 14.5 degrees, an observation of 5-minutes duration would be executed. We can follow the MWA strategy of observing around the Sun, which would require a total of 14 such pointings around the Sun at this frequency as shown in Figure \ref{fig:observing_grid}, thus covering a total of 770 sq degree of sky around the Sun in 70 minutes. 

Since the Sun moves $\sim$1 degree across the sky in one day, this strategy would require observing the sky only once every 7 days to optimally cover the sky to conduct an IPS survey. Therefore, the observing time required is only 70 minutes per week at this frequency.

Additionally, these observations are made in the daytime. Based on experiences with MWA and ASKAP, these are times when the telescopes are least utilised to avoid the unwanted effects of the Sun on other surveys. Thus, these short observations can be conducted as filler times between observations increasing the usage of the SKA-Low. 

 


\subsection{Density of sources detectable by the SKA-Low}
The MWA IPS studies provide the most useful basis to compare sensitivity and estimate source densities for SKA-Low. In Table \ref{Tab:sensitivity_variability_image} we estimate the 5-sigma IPS sensitivity  based on 1-sigma continuum image sensitivity achieved by SKA-Low using the inner 6-km stations at 162\,MHz. The 1-sigma continuum image sensitivity was calculated using the SKA sensitivity calculator using Natural weighting scheme. \cite{Morgan2017} found that the sensitivity of the variability images only increases as the fourth root of total integration time. Table \ref{Tab:sensitivity_variability_image} shows that 5 minute of total integration time provides sufficient sensitivity to probe compact sources at $\sim$ 1\,mJy level, while ensuring sufficient time stamps are obtained to measure IPS properly. This is expected to be the default observing time for IPS observations in imaging mode.

\begin{table}[ht]
	\centering
	\caption{Estimated 5-sigma sensitivity achieved to detect sub-arcsecond scale objects at 162\,MHz using 32\,MHz bandwidth in images made using only the inner 6-km stations of AA4. Natural weighting scheme has been assumed for imaging. 1-$\sigma$ continuum sensitivity was estimated using SKA sensitivity calculator.}
	\label{Tab:sensitivity_variability_image}
	\begin{tabular}{ccccc}
			\hline
			Time intervals & 1-$\sigma$ continuum sensitivity & \multicolumn{3}{c}{5-$\sigma$ IPS sensitivity (mJy/beam) for observations of:}\\
			
			(seconds) & (mJy/beam) & 1 minute  & 2 minutes  & 5 minutes\\
			\hline
			0.28 & 1.67 & 1.38 & 1.16 & 0.92\\
			\hline
		\end{tabular}
\end{table}

\begin{figure}[ht]
	\centering
	\includegraphics[width=0.8\columnwidth]{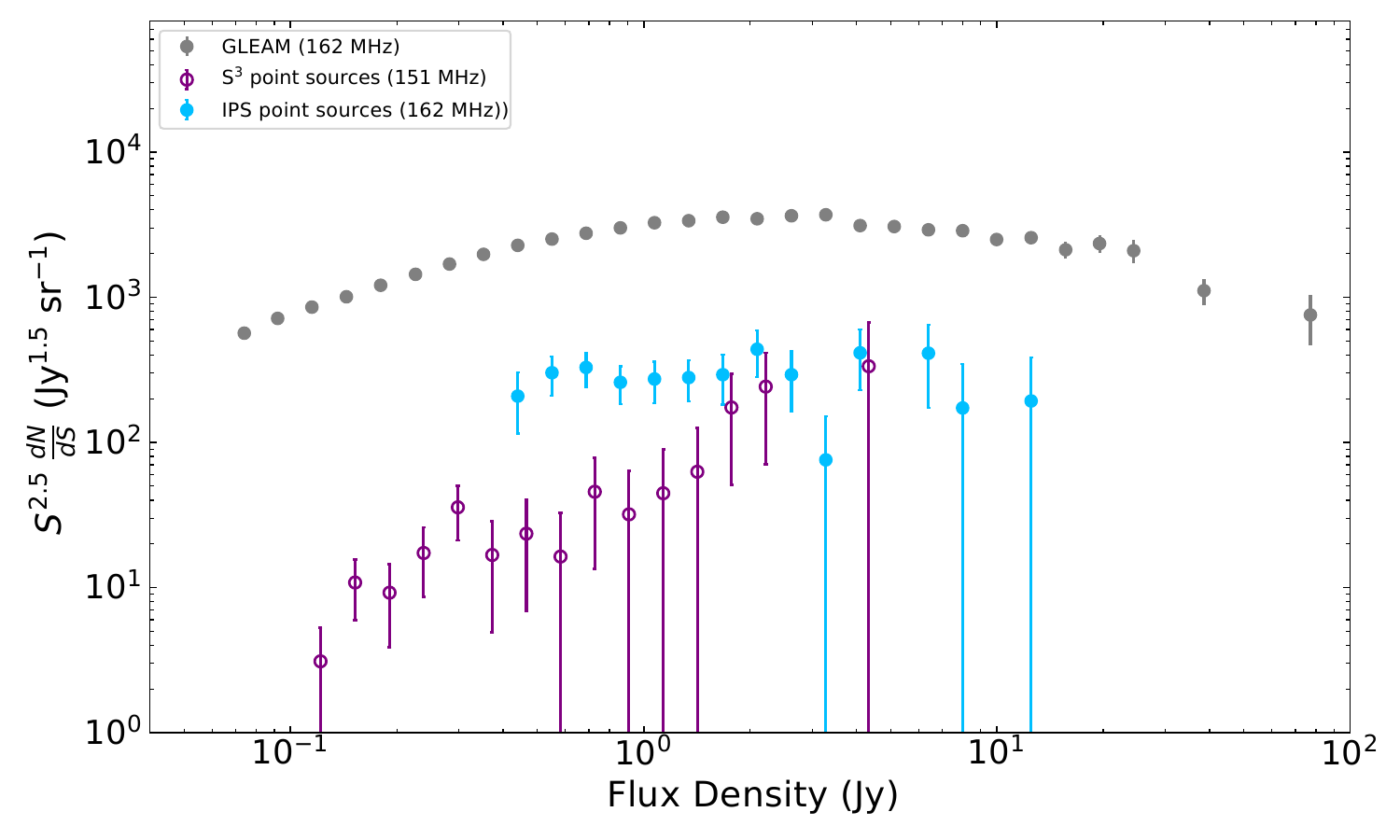}
	\caption{Source count of sub-arcsecond sources detected by the MWA compared with that expected from the S$^3$ simulations (from \cite{Chhetri2018-IPS3}). The grey dots are the source counts from GLEAM survey.}
	\label{fig:ppr2_src_count}
\end{figure}

\citet{Chhetri2018-IPS3} made source counts of point sources using single wide-field observation of IPS with the MWA. They define "point sources" as objects whose flux density arises entirely from sub-arcsecond scale compact regions (e.g. AGN cores). They compared the results of their detections of compact sources with that expected from the SKA Simulated Skies \citep[S$^3$,][]{S3-Wilman2008}. This comparison is reproduced here as Figure \ref{fig:ppr2_src_count}. The counts of observed sources shows that the number density of sources is much higher than expected from simulations at fainter flux densities. 

By extrapolating this measured source count using a linear fit down to the sensitivity level we expect to reach (Table \ref{Tab:sensitivity_variability_image}), we estimate that SKA-Low can expect to detect sources at the density of 409 sources per square degrees at 2\,mJy/beam sensitivity level. We have used this conservative estimate of sensitivity to allow for drop in sensitivity possible due to different reasons such as unavailability of all stations, solar interference etc. At this sensitivity level, we expect the standard continuum image to be source confusion limited. However, our image-based approach of detecting scintillating sources in variance image means that only scintillating sources remain in the image. Therefore, with 409 sources/square deg., we estimate $\sim$\,15 synthesized beam/source\footnote{The SKA sensitivity calculator currently does not provide synthesized beam sizes for Natural weighting. Therefore, to make this estimate we have used a conservative value compared to Uniform weighting.} indicating that source confusion will not limit our detections. 
\textit{We note here that this will be achieved with only five minute of observations, therefore highlighting an efficiency which far exceeds that of traditional VLBI approach to identify compact objects at sub-arcsecond scale. }

\subsubsection{Impact of detection of high density of compact sources}
This high source density of sources will enable studies of compact source populations indicated in the first part of this chapter to unprecedented sensitivity level. 

\begin{figure}[ht]
	\centering
	\includegraphics[width=0.9\columnwidth]{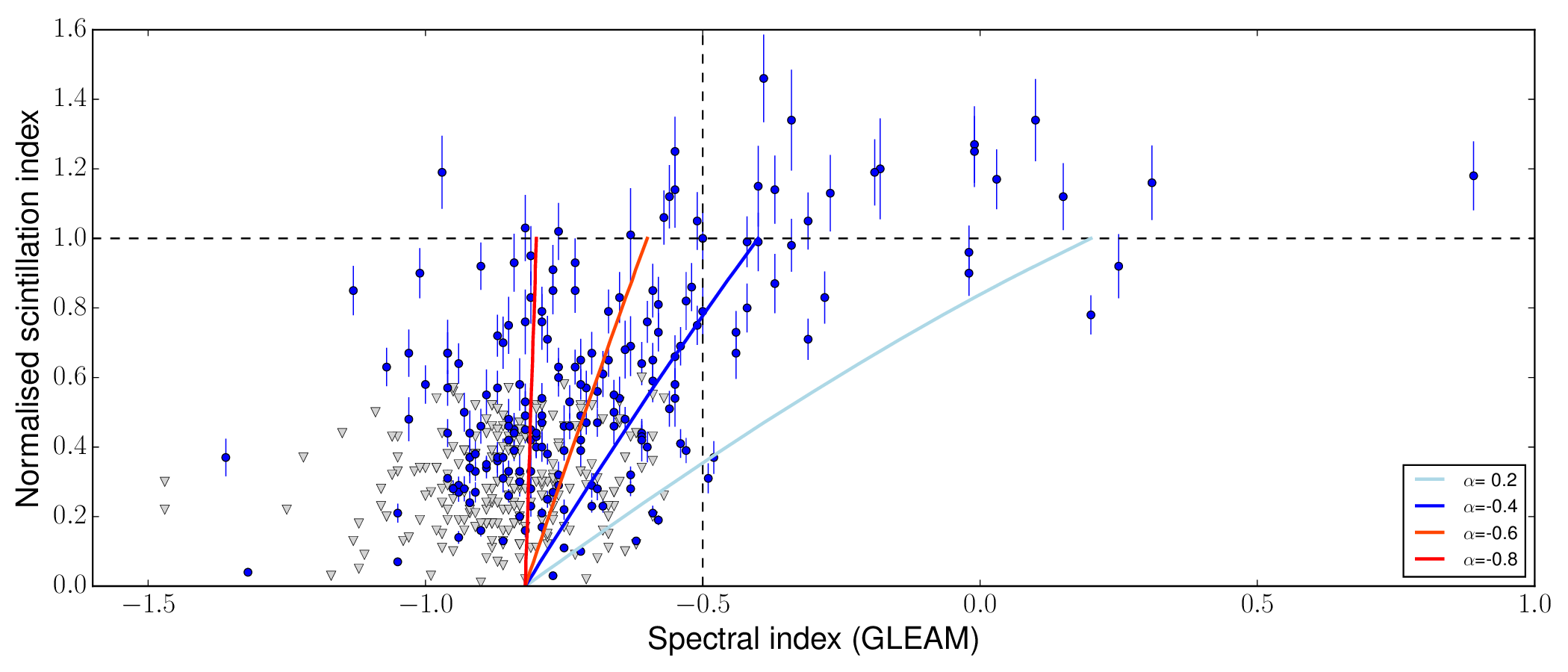}
	\caption{Distribution of NSI (determined by source angular size) as a function of spectral index reproduced from \cite{Chhetri2018-IPS2}. Sources size increases with lowering NSI. The coloured lines indicate spectral indices of compact components that is embedded in extended structures. Blue dots are based on measurements and grey dots are for upper limits for GLEAM sources in the field of view. }
	\label{fig:ppr2_alpha_tracks}
\end{figure}

Figure \ref{fig:ppr2_alpha_tracks} (reproduced from \cite{Chhetri2018-IPS2}) is a snapshot of the different radio source populations based on angular size (represented by normalised scintillation index (NSI)) and spectral index detected by radio surveys. This plot provides important insight into the relationship between the compact cores and their extended radio galaxy components (jets and lobes).  It also indicates the pathways they may take from nascent compact cores to extended radio galaxies and to remnant phases of radio galaxies. As we probe different sensitivity levels with SKA-Low, it will enable us to understand how the different source populations evolve relative to each other. Such plots also provide important diagnostics for new and unexpected objects, and source-populations as they are serendipitously detected by SKA-Low.   

Additionally, at this flux density level, traditional surveys will identify extremely large number of objects, many of which will be star-forming galaxies. One important challenge that will be faced by these studies is to delineate the AGN contribution in these star-forming galaxies. Brightness temperature has been widely used to identify AGNs in the past, and more recently with LOFAR in the low frequency regime \citep{morabito_2022MNRAS.515.5758M}. Due to the compact nature of the high brightness temperature AGN cores, they are expected to be scintillating in distant galaxies but the relatively low brightness temperature star forming regions won't. \textit{Thus, the mJy level IPS sensitivity reached using SKA-Low will provide a very efficient approach to distinguish AGN contributions in star forming galaxies.} 

We once again note that the high density of compact sources identified by IPS technique will provide the \textit{network of calibrators for the SKA-Low} required to make direction-dependent ionospheric calibration.

Furthermore, this high density of compact sources will form the basis to detect and \textit{study space weather events such as coronal mass ejections (CMEs) in very high definition} using an approach recently developed with the MWA \citep{Waszewski2023}. 
This and other impacts on heliospheric studies of this high density detection of sources will be the focus of the remainder of this chapter.  

\subsection{Widefield Imaging IPS for Space Weather monitoring}
\label{subsec:ska_imaging_swx}
The approach outlined in the previous section has also been used successfully with the MWA for Space Weather studies.
In the particular case of the MWA, with its extremely wide field of view, the same observations were used for both space weather monitoring and for astrophysical studies.
With a small number of pointings it was possible to cover the full accessible annulus around the Sun \citep[e.g.][]{2025SpWea..2304342W}.
For the SKA, this may not be the best approach, since the field of view is considerably smaller and so multiple observations (or multiple beams) will be needed in order to cover the same fields covered by the MWA in a smaller number of pointings.

Given that IPS sensitivity is not easily increased by observing for longer (detection threshold only goes down with the fourth root of observing power) observing with the full collecting area will always be more efficient for surveys.
However, in order to cover a wider field of view as rapidly as possible, another approach would be to observe with a sub-array using sub-stations \citep[e.g. using template \texttt{Low\_Substation\_18m}][]{Trott2024};  the sky coverage would be substantially increased, although with a correspondingly lowered sensitivity.

Additionally, in unpublished work, it has been shown that with a beamformer which can set the complex gain for each antenna of an SKA-Low station, it is possible to null the Sun while maximising the sensitivity of a nearby field, or even a produce a primary beam shape forming an annulus around the Sun. 

\subsection{Multi-station IPS with the SKAO telescopes}
\label{subsec:ska_multi}
As discussed in Section~\ref{subsubsec:swx_multistation}, IPS space weather observations to-date have principally been conducted using a ``multi-station'' approach where the scintillation pattern is measured at geographically dispersed locations.
What distinguishes SKA-Low from previous instruments is not only its immense collecting area and sensitivity, but its distributed aperture-array design, in which hundreds of stations are arranged in a dense central core with spiral arms extending tens of kilometres.
There is therefore the potential to utilise SKA-Low as a powerful multi-baseline IPS instrument, provided that it can be operated in a mode where independent tied-array beams are formed at selected station clusters.
By beamforming a small number of compact background sources at these stations, especially those located at the ends of the arms, one can perform simultaneous measurements of IPS fluctuations from widely separated locations.

The wealth of information that can be gleaned from a large number of baselines has been demonstrated by International LOFAR \citep{2023AdSpR..72.5311F}.
However, the baselines of the SKA-Low are much shorter than those of international LOFAR.
Therefore, before considering the technical feasibility of this approach with SKA-low, it is important to carefully assess whether the baselines of SKA-Low are of a suitable length.

The baselines between the outermost clusters on the three SKA arms are comparable with the baselines used in early studies of multi-station IPS \citep{1966AJ.....71Q.854D,1967Natur.213..343D}, and are just a little shorter than the purpose-built UCSD and ISEE arrays discussed in Section~\ref{subsubsec:swx_multistation}.
Additionally, \citet{1971A&A....10..310E}, working at 2.3\,GHz, where the spatial scale of the scintillation pattern is much smaller, were able to measure the velocity of the solar wind (and its acceleration via observations at different solar elongations), with an antenna separation of only 16\,km.
There is therefore little doubt that velocity measurements could be made with SKA-Low provided that it is possible to generate timeseries with the requisite time resolution for the clusters at the ends of each of the spiral arms.

However, \citet{1996Ap&SS.243...87C} stress the importance of longer baselines to separate multiple streams along the line of sight.
\citet{1996Natur.379..429G}, with an observing frequency of 933\,MHz, were only able to fully resolve fast and slow streams with a baseline of 240\,km.
Therefore, there will most certainly be interest in having access to longer baselines than the AA4 SKA-Low configuration envisages.
One potential route to this is The Low-frequency Australian Megametre-Baseline Demonstrator Array \citep[LAMBDA;][]{2024evn..conf..189R}.
LAMBDA is a pioneering concept for a small number of low-frequency stations spread over large baselines and using signal-processing techniques compatible with SKA-Low.
For IPS, these would ideally span baseline lengths similar to International LOFAR: from $\sim$100\,km to $\sim$1000\,km.
IPS measurements across these lengths have the potential to reveal the anisotropy of scattering, departures from radial outflow, variations in the orientation of turbulence, and the coherence length of density structures in multiple streams along the Line of sight.

Another relevant consideration is whether a shift to SKA-Mid might be a better choice in some cases.
While SKA-Low operates across roughly 50\,MHz--350\,MHz, SKA-Mid covers significantly higher frequencies, where observations closer to the Sun provide a better match.
Crucially for a multi-station approach, the scintillation pattern is also finer at higher frequencies \citep[the spatial scale of the pattern scales with $\sim\sqrt{\lambda}$;][]{1992RSPTA.341..151N}.
Therefore, along with the ability of SKA-Mid to observe closer to the Sun, it may also permit the resolution of different streams along the line of sight in the cross-correlation function.

Finally, there are two untested approaches that may provide additional information and/or allow for different data products to be used for IPS.
Firstly, while weak scatter is dominated by structure on the Fresnel scale, there are also weaker signals at higher frequencies due to higher-order terms.
Due to the negative power-law structure function of this turbulence, this signal is much weaker.
However, if the high sensitivity of the core of SKA-Low can be utilised, there may be sufficient signal-to-noise to filter out this component.
This would result in a much narrower autocorrelation function, and this should allow closely spaced peaks from different streams in the cross-correlation function to be resolved.

Another alternative approach worthy of consideration is the ``visibility scintillation'' approach of \cite{1972ApJ...174..181C}.
Standard IPS techniques rely on temporal intensity fluctuations, but interferometric visibilities contain both amplitude and phase (formally, the difference in phase between antennas of a baseline).
The phase term is particularly rich in physical information: it encodes the velocity vector of the scattering screen as well as refractive wander, large-scale bending, and the ensemble-average scatter-broadening of point sources.
For many instruments, measurements from a single baseline are too noisy for the phase to be tracked.
However, \citet{2025PASA...42...55J} were able to exploit very wide bandwidth VLBI to track the temporal evolution of phase fluctuations caused by the solar wind for a small number of baselines.
If this is possible with the SKA instruments for brighter sources,
the full visibility information \emph{and} its change with time over the full aperture plane is captured. At this point there is no distinction between scintillation and scatter broadening (discussed in detail by Zhang et al. in this volume).
How and whether the full information encoded in this dataset can be extracted will require future dedicated studies.

To conclude this section, we note that while multi-station IPS observables have been shown to provide strong constraints on the turbulent structures in the solar wind \citep[see][]{2005JGRA..110.3101H}, analyses even of a small number of cross-correlation functions is non-trivial, since we are dealing with a diffraction pattern arising from multiple phase screens along the line of sight, which is evolving spatially and temporally \citet{1971A&A....10..306L}.
The improvement in sensitivity and the unprecedented wide bandwidth of SKA-Low will provide extremely detailed information at very high signal-to-noise, and new tools may need to be developed in order to optimise the science output.

\subsection{Synergies with other instruments}
Interplanetary scintillation (IPS) is an important tool for remotely studying the density and velocity structure of the inner heliosphere (roughly 0.1--1.0 AU) \citep{Hewish1964, Little1966a, Little1966b, Bisoi2014b, Tanti2025}. 
Long-term, multi-station IPS datasets have been used to map large-scale solar wind streams, detect transient structures (CMEs, shocks), and quantify mesoscale turbulence levels, information that is difficult or impossible to obtain from single-point in-situ probes alone \citep{Nindos_2019, 2023AdSpR..72.5341J}. 
The coming decade will see a step-change in corona and heliosphere monitoring through the combination of next-generation radio arrays (notably SKA-LOW/MID), modern wide-field low-frequency facilities (LOFAR, LWA) and expanding national IPS networks (ISEE, Mingantu/MUSER, FAST/China), coupled with the continuous in-situ sampling of Parker Solar Probe (PSP), Solar Orbiter and the L1 monitors like India's Aditya L1 \citep{Tokumaru2011, Bisoi2014b, Nindos_2019, 2023AdSpR..72.5341J, Tanti2025}.

Dedicated multi-station IPS observatories like the ISEE network with 327-MHz cylindrical/parabolic arrays provide multi-baseline time-delay measurements that directly yield line-of-sight solar-wind speeds and microstructure diagnostics via cross-correlation techniques \citep{Tokumaru2011, Bisoi2014b, Ingale2019, Tanti2025}.
These dedicated systems will remain essential for regular monitoring (Since the SKA is an all-purpose instrument, it is likely to be used only for observing ``campaigns'', and to exploit targets of opportunity);
providing robust velocity and density daily, and providing input to tomographic inversions derived from wide-field arrays \citep{Tokumaru2011, Tokumaru2023}. 
The global Worldwide IPS Stations \citep[WIPSS;][]{2023AdSpR..72.5341J,2023AdSpR..72.5328I} concept shows how combining multi-site IPS streams improves CME arrival predictions and 3D reconstruction quality.
The upcoming SKAO telescopes (Low and Mid) will bring both orders-of-magnitude sensitivity increases, and flexible beamforming/imaging modes to IPS science \citep{Nindos_2019}. 
SKA-Low's dense aperture arrays are ideal for high-cadence, broad-sky scintillation surveys (extending current MWA work to much greater sensitivity), while SKA-Mid dishes can operate at higher frequencies closer to the Sun, and can link more directly with ground- and spaced-based coronagraph/white-light observations. 

This will allow development of a synergetic pipeline which is tomographic inversion constrained by heterogeneous IPS inputs: wide-field scintillation indices (initially from MWA, then SKA-Low) supply dense angular sampling of the scintillation field; multi-baseline time-delay data from LOFAR and ISEE-style arrays fix the local speed along particular lines of sight; and modern IPS arrays under construction (e.g., Mingantu/MUSER and Chinese cylinder arrays) will add intermediate baselines and frequency coverage to sample different turbulence scales \citep{Waszewski2023}.
Combining these measurements in a unified inversion framework (kinematic tomography with data assimilation into MHD models) will substantially improve 3-D CME/corpus reconstructions, yield better CME mass and speed estimates, and reduce arrival-time uncertainties:  critical for space-weather forecasting \citep{Waszewski2023}. 
Recent studies demonstrate the power of adding wide-field IPS data into heliospheric reconstructions to resolve moving features not captured by sparse vantage points alone \citep{Waszewski2023}.
Moreover, different instruments and frequencies probe different scales of density inhomogeneity. 

Interplanetary Scintillation observations have long been combined with white-light images from space-based platforms, such as the Solar Mass Ejection Imager \citep[SMEI;][]{2003SoPh..217..319E, 2020FrASS...7...76J}.
Other Heliospheric Imagers provide complementary fields of view \citep{2020SoPh..295..150B}, and a new generation of instruments with wider fields of view, such as the Polarimeter to UNify the Corona and Heliosphere \citep[PUNCH;][]{2022aero.confE...1D} there will be ample opportunities to explore the synergies and complementarity of polarised Thompson scattering with IPS.

Higher-frequency IPS can operate much closer to the Sun, overlapping with space-based coronagraph.

By exploiting multi-frequency, multi-baseline observations one can disentangle transient large-scale density structures (CMEs, SIRs) from the ambient turbulent cascade and quantify how turbulence amplitude and spectral slope evolve with heliocentric distance and in the presence of transients. 
This multiscale view is critical to separating propagation effects (e.g., scattering) from source effects (e.g., density enhancements inside CMEs) and to testing turbulence transport models used in space-weather prediction systems \citep{Morgan2017, Morgan2019}.

Additionally, Spacecraft such as Parker Solar Probe and Solar Orbiter are providing unprecedented in-situ measurements of fields and particles deep in the corona and inner heliosphere \citep{2023AdSpR..72.5341J}. 
However, they are single or few point samplers. IPS provides the synoptic spatial context that places in-situ encounters into a global, time-evolving heliospheric picture. 
For example: (a) IPS ``g-maps'' can show a CME's large-scale density front and speed profile at the time PSP or Solar Orbiter samples a local patch, enabling direct comparison between remote density/velocity structure and local microphysics measurements; (b) IPS-derived turbulence measures across many lines of sight can be compared with PSP turbulence spectra, testing models of turbulence evolution (cascade, anisotropy, intermittency) as the wind expands. 
Cross-validation of remote and in-situ diagnostics is essential to constrain theories of wave-driven heating and the formation of small-scale structures that seed large-scale variability \citep{Velli2020}.

\section{Summary}
\label{sec:summary}

We have shown that using the technique of interplanetary scintillation in the wide field-of-view of SKA-Low provides an opportunity to identify sub-arcsecond scale sources in large numbers with very modest investment of time with the SKA-Low and that it will enable unprecedented research in the study of compact source population and space weather. We outlined a potential approach to conducting such an IPS survey with SKA-Low, and estimated that such a study will provide a compact source density of 409 sources/square degree at 2\,mJy/beam level, providing the grid of compact sources important for SKA-Low to make direction dependent calibration of multiple ionospheric patches in SKA-Low images. We further outlined a wide range of astrophysics problems that such a survey would positively impact such as the search for pulsars, high redshift radio galaxies, and HI absorption.  

Additionally, IPS observations can also be made in order to support research into the Heliosphere, as well as space weather monitoring.
Such observations could be done in an imaging mode (similar to the way that observations are made with the MWA currently), or utilise a beamforming approach using the outer stations, possibly also with a phased core.
This would enable observations similar to those conducted by LOFAR, although longer baselines than those envisaged for the AA4 layout would be highly desirable.
\bibliographystyle{abbrvnat-maxbibnames4}
\bibliography{chapter}

\end{document}

%% file: journal-names.tex
\newcommand{\actaa}{Acta Astron.} 
\newcommand{\araa}{ARA\&A} 
\newcommand{\aar}{A\&ARv} 
\newcommand{\aapr}{A\&ARv} 
\newcommand{\ab}{Astrobiol.} 
\newcommand{\aj}{AJ} 
\newcommand{\apj}{ApJ} 
\newcommand{\apjl}{ApJL} 
\newcommand{\apjs}{ApJSS} 
\newcommand{\ao}{Appl. Opt.} 
\newcommand{\apss}{Astro. \& Space Sci.} 
\newcommand{\aap}{A\&A} 
\newcommand{\aaps}{A\&AS.} 
\newcommand{\baas}{Bull. Am. Astron. Soc.} 
\newcommand{\caa}{Chinese A\&A} 
\newcommand{\cjaa}{Chinese J. A\&A} 
\newcommand{\cqg}{Class. Quantum Gravity} 
\newcommand{\gal}{Galaxies} 
\newcommand{\gca}{Geo. Cosmo. Acta} 
\newcommand{\icarus}{Icarus} 
\newcommand{\jcap}{JCAP} 
\newcommand{\jgr}{J. Geophys. Res.} 
\newcommand{\jgrp}{J. Geophys. Res. Planets} 
\newcommand{\jqsrt}{J. Quant. Spectrosc. Radiat. Transf.} 
\newcommand{\memsai}{Mem. SAIt} 
\newcommand{\mnras}{MNRAS} 
\newcommand{\nat}{Nature} 
\newcommand{\nastro}{Nat. Astron.} 
\newcommand{\ncomms}{Nat. Commun.} 
\newcommand{\nphys}{Nat. Phys.} 
\newcommand{\na}{New Astron.} 
\newcommand{\nar}{New Astron. Rev.} 
\newcommand{\physrep}{Phys. Rep.} 
\newcommand{\pra}{Phys. Rev. A} 
\newcommand{\prb}{Phys. Rev. B} 
\newcommand{\prc}{Phys. Rev. C} 
\newcommand{\prd}{Phys. Rev. D} 
\newcommand{\pre}{Phys. Rev. E} 
\newcommand{\prx}{Phys. Rev. X} 
\newcommand{\prl}{Phys. Rev. Let.} 
\newcommand{\psj}{Planet. Sci. J.} 
\newcommand{\planss}{Planet. Space Sci.} 
\newcommand{\pnas}{Proc. Natl Acad. Sci. USA} 
\newcommand{\procspie}{Proc. SPIE} 
\newcommand{\pasa}{PASA} 
\newcommand{\pasj}{PASJ} 
\newcommand{\pasp}{PASP} 
\newcommand{\rmxaa}{RMXAA} 
\newcommand{\sci}{Science} 
\newcommand{\sciadv}{Sci. Adv.} 
\newcommand{\solphys}{Sol. Phys.} 
\newcommand{\sovast}{Soviet Ast.} 
\newcommand{\ssr}{Space Sci. Rev.} 
\newcommand{\uni}{Universe} 